\documentclass[epj]{svjour}
\usepackage{latexsym}
\usepackage{graphics}
\usepackage{color}
\usepackage{amsmath}
\usepackage{amssymb}
\usepackage{graphicx}
\usepackage{wasysym}
\usepackage{ulem}
 
\begin{document}
\title{Let's deflate that beach ball}
\author{Gwennou Coupier\inst{1} \and Adel Djellouli \inst{1}
\thanks{\emph{Present address:} Harvard John A. Paulson School of Engineering and Applied Sciences, Harvard University, Cambridge, Massachusetts 02138, USA} \and Catherine Quilliet  \inst{1}
}                     % Do not remove
\institute{Univ. Grenoble Alpes, CNRS, LIPhy, 38000 Grenoble, France}
\date{Received: date / Revised version: date}
\abstract{We investigate the relationship between pre-buckling and post-buckling states as a function of shell properties, within the deflation process of shells of an isotropic material. With an original and low-cost set-up that allows to measure simultaneously volume and pressure, elastic shells whose relative thicknesses span  on a broad range are deflated until they buckle. We characterize the post-buckling state in the pressure-volume diagram, but also the relaxation toward this state. The main result is that before as well as after the buckling, the shells behave in a way compatible with predictions generated  through thin shell assumption, and that this consistency persists for shells where the thickness reaches up to 0.3 the shell's midsurface radius. 
} %end of abstract
\maketitle
\section{Introduction}

Due to the boom of microfluidics and miniaturization, small spherical
objects are increasingly studied in soft matter, many
of them thin and prone to deformation. Deformation is usually accompanied by deflation (\textsl{e.g.} due to osmotic pressure,
or leakage, or lateral expansion of the shell). There have been several
theoretical or numerical studies \cite{Hutchinson 1967,LandauBook,Quilliet2008,Quilliet2008err,Knoche2011,Vliegen2011,Quilliet_2012,Hutchinson_2017,pezzulla2018} and some experimental investigations \cite{Carlson_1967,Carlson_1968,Zhang_2017}
 about the deflation of a thin, elastic,
shell. Most of them focus essentially on understanding and quantifying  the scenario of the buckling instability that occurs beyond a certain threshold of compression or deflation. Less is known about the post-buckling behaviour  \cite{Quilliet2008,Quilliet2008err,Knoche2011,Quilliet_2012,Knoche2014}, let alone when thin shell theory is \textit{a  priori} not valid. It is generally assumed that a 2D description of the shell is valid when $d/R < 0.02$, where $d$ is the shell thickness and $R$ its mid-surface radius ($R-\frac{d}{2}$
and $R+\frac{d}{2}$ are then respectively the internal and external radii). In that case the  2D properties of the surface model
can be interpreted in terms of shell thickness and 3D properties
of the constituting material. These models  indeed
constitute a simplification compared to studies managing 3D features
\cite{Church_1994,Knoche2011}.

\begin{figure}
\resizebox{\columnwidth}{!}{\includegraphics{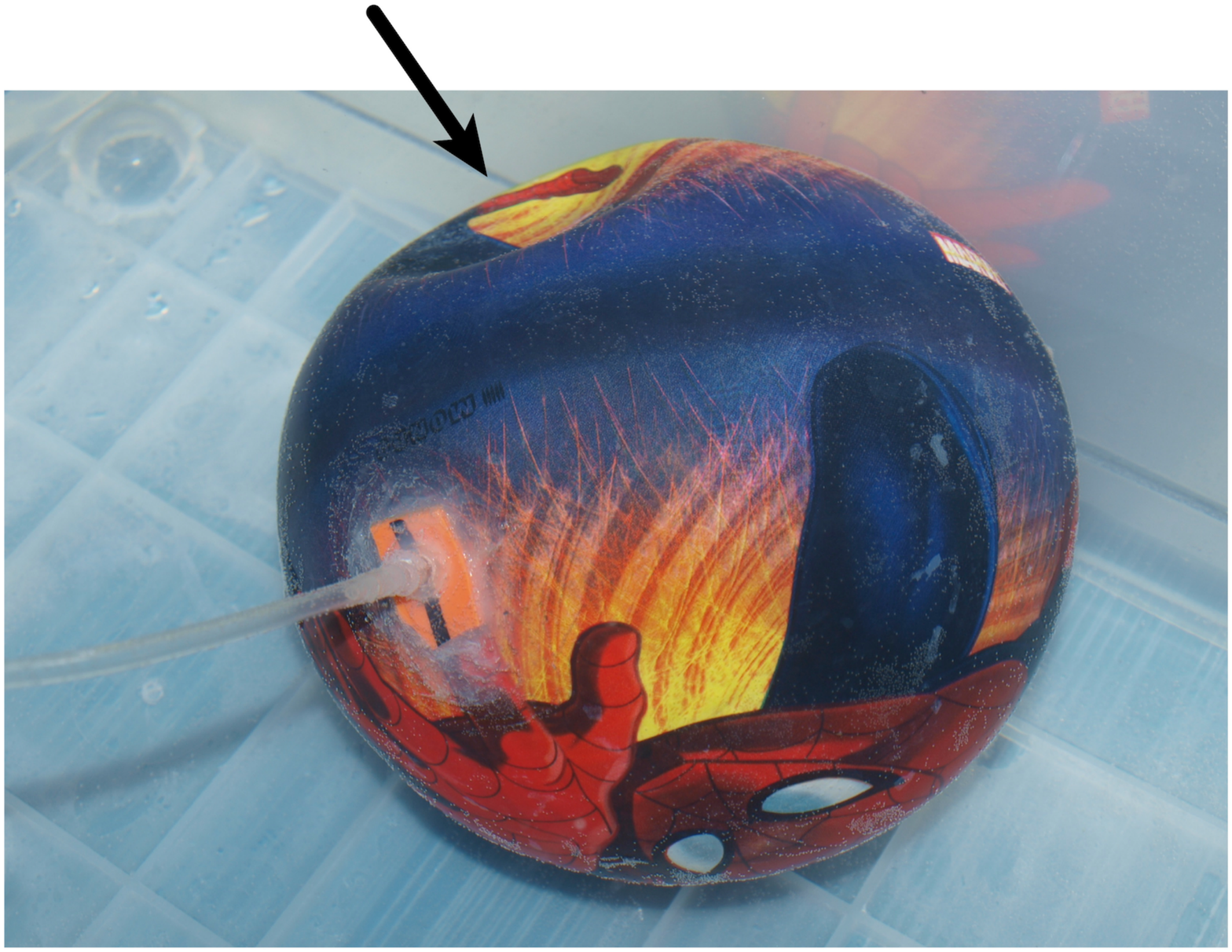}}
\caption{\label{fig:spi}Buckled shell through volume loss: an axisymmetric
depression (shown by arrow) suddenly appears  when internal volume
is slowly decreased. External radius $R+d/2=171$ mm; relative thickness
$d/R=6.5\thinspace10^{-3}$.}
\end{figure}

In this paper, we investigate experimentally the deflation of elastic macroscopic shells, down to buckling
and post-buckling deformations, for a broad range of relative shell thicknesses. These results are compared to what is known from thin shell theory, which allows to discuss its validity range.

Our low-cost experimental set-up was conceived as an efficient and versatile tool for exploring with students instability issues and bifurcations diagrams under several conditions (volume or pressure imposed), and for characterizing shells before using them in a more complex environment \cite{Djellouli_2017}. Yet, it provides for the first time an experimental characterization of the relationship between pre-buckling and post-buckling states. Transition between these two states is accompanied by a fast release of energy, a feature present in Nature \cite{forterre05,vincent11,son2013} that has already been used in several applications with similar soft systems \cite{Djellouli_2017,holmes07,yang15,ramachandran2016,gomez2017,holmes2019}.

\begin{figure}
\resizebox{\columnwidth}{!}{\includegraphics{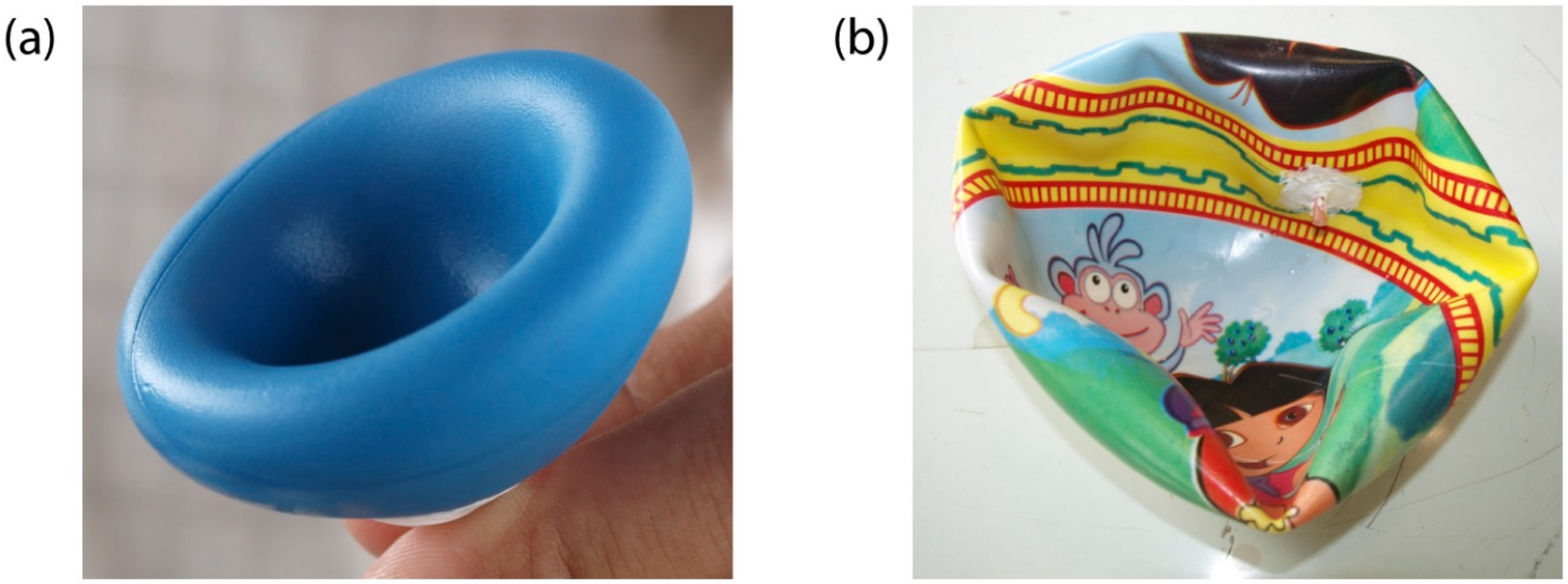}}
\caption{\label{fig:photo-balles}Elastic
shells almost fully deflated; the hemisphere that is curved inward
is in contact with the, mainly unchanged, other half of the shell.
(a) Thick shell (relative thickness $d/R=0.18$; external radius $R+d/2=55.5\,\mathrm{mm}$),
in which axisymmetry was conserved throughout the deflation. (b) Thin
shell ($d/R=10.6\thinspace10^{-3}$; external radius $R+d/2=190\,\mathrm{mm}$),
where radial folds began to develop inside the depression at some
point of the deflation process.}
\end{figure}

Spherical shells of an isotropic elastic material are expected to undergo
sequences of shapes that depend only on intensive parameters: Poisson's
ratio $\nu$ and relative thickness $d/R$.
Shells should first stay spherical while their radius decreases, up
to the point where a buckling instability suddenly makes  a
circular depression appear, of characteristic dimension $\sqrt{dR}$ (Fig.
\ref{fig:spi}) \cite{LandauBook,Pogorelov}. This step was only recently
understood in terms of mode localization \cite{Hutchinson_2017,Hutchinson_2016}.
According to simulations and theoretical studies, the depression then
grows axisymmetrically  when the shell is slowly deflated 
\cite{Quilliet2008,Quilliet2008err,Knoche2011,Hutchinson_2017}. Thicker shells keep
axisymmetry up to self-contact (Fig. \ref{fig:photo-balles}-a), while
for thinner shells, the depression looses its axisymmetry
during deflation, progressively developing 
radial folds (Fig. \ref{fig:photo-balles}-b) \cite{Quilliet2008,Quilliet2008err,Quilliet_2012,Hutchinson_2017,Knoche2014}.

Quantitatively, the deflation is characterized by the volume change
$\Delta V$ from the initial nondeflated state, and the pressure drop
$\Delta P=P_{ext}-P_{int}>0$ it induces between both sides of the shell
(outside and inside the ball). In a surface model, the denominator
of the dimensionless relative volume variation $\frac{\Delta V}{V_0}$
is the volume enclosed by the initial undeformed surface. For
this experimental study we chose to take as a reference the volume
$V_{0}=\frac{4}{3}\pi R^{3}$ initially enclosed by the midsurface
of the shell, instead of the volume $V_{int}=\frac{4}{3}\pi\left(R-\frac{d}{2}\right)^{3}$
effectively contained in the shell, thus allowing direct comparison with surface models. The set-up we developed provides the pressure drop and the
volume variation of deflated spheres of known initial volume ; we
could then follow and discuss deformation paths observed in a $\Delta P\,-\,\frac{\Delta V}{V_0}$
diagram. We denote by $\wp(\frac{\Delta V}{V_0})$ the state equation between
both quantities at equilibrium. This function $\wp$  is to be determined in this paper.

\begin{figure*}
\begin{center}
\resizebox{1.8\columnwidth}{!}{\includegraphics{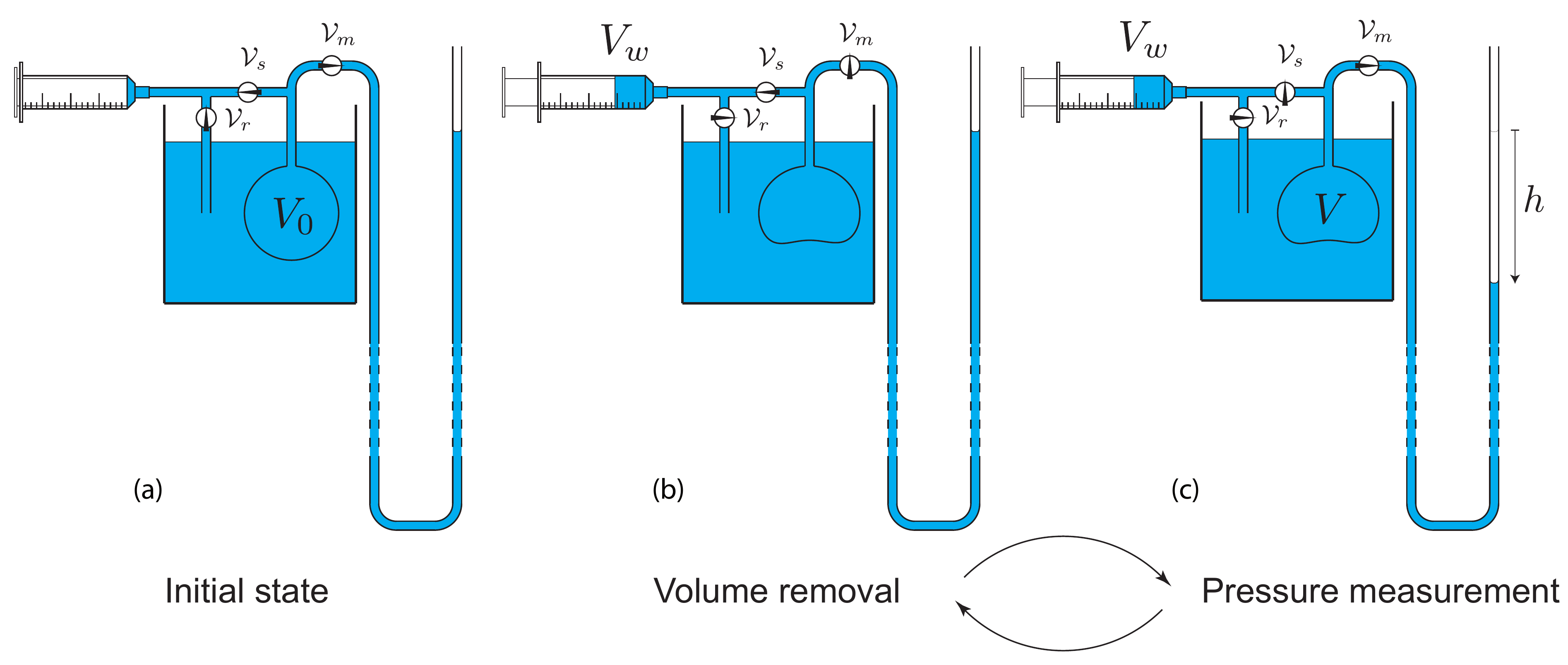}}
\caption{\label{fig:Principle} Principle of the experiment: an elastomer ball
is plunged in a tank of water. The ball is filled with the same liquid,
which avoids extra deformations due to internal-external differences
of hydrostatic pressure. Its inner volume is linked either to a manometer
(U tube on the right) through valve $\mathcal{V}_{m}$, or, through
valve $\mathcal{V}_{s}$, to a syringe that modifies the volume whatever
 the pressure, or to the water tank through valves $\mathcal{V}_{s}$
and $\mathcal{V}_{r}$. (a) Initial state: pressures in the ball,
in the manometer and in the water tank are equilibrated ($\mathcal{V}_{m}$,
$\mathcal{V}_{s}$ and $\mathcal{V}_{r}$ open). The initial volume
enclosed by its midsurface is $V_{0}=\frac{4}{3}\pi R^{3}$ (reference
for subsequent calculation of the relative volume variation). At rest,
the level in the manometer (which is a tiny tube) overtakes that in the tank by about a few millimeters, due to capillary
effects. This level at rest is taken as the reference for pressure
calculations. (b) Deflation: through open $\mathcal{V}_{s}$, the
syringe removes a controlled volume of water $\delta V_{i}$ (this
volume can safely be considered as incompressible even if inner pressure
drastically drops). (c) After the closing of $\mathcal{V}_{s}$, the opening
of $\mathcal{V}_{m}$, and equilibration, the pressure difference between
both sides of the ball wall is calculated from the slump $h>0$ of
water level in the manometer:
$\Delta P=\rho gh$. The inner volume change due to the level adjustment
in the manometer is not negligible and is taken into account in further
calculations. Practically, deflation is done stepwise, through repetition
of stages (b) and (c), this latter providing a series of equilibrium
states corresponding to inner volume variations $\left(\Delta V_{1}, ..., \Delta V_{i},\Delta V_{i+1}, ...\right)$.
When full, the syringe can be emptied through valve $\mathcal{V}_{r}$
without opening the system.}\end{center}
\end{figure*}

\section{Set-up for deflation experiments}

We considered about 25 commercial hollow balls (beach balls, squash balls,
juggle balls, balls for rhythmic gymnastics...) made of elastomers, of external radii $R+d/2$ ranging
between 39.5 and 190 mm, and $d/R$ ratios between $6.5\thinspace10^{-3}$
and 0.25, plus a homemade ball of relative thickness 0.22 \cite{Djellouli_2017}. All Young moduli $Y_{3D}$ measured for small strains are between 0.5 and  7.5 MPa (see  section \ref{sec:Traction}).

In order to easily measure volumes and pressures, the ball is filled
with an incompressible fluid (water). It is then also immersed in
water so as to avoid gradients of hydrostatic pressure along the ball
(which amounts to study shapes not deformed by gravity). The ball
is connected to a U-shape manometer,  a syringe,
and a third tube connected to the tank of water so as to favor initial quick
equilibration of all pressures (Fig. \ref{fig:Principle}-a). In the
initial state, the pressure difference $\Delta P$ is 0. Taps allow us
to connect the ball either to the syringe or to the manometer.  The manometer is made of a cylindrical tube of diameter ranging between 0.79 and 3.18 mm and thick enough to avoid tube buckling
under the highest pressure differences 1 bar, which are met with  thick squash balls. If required, the total height of the manometer could reach $10\,\mathrm{m}$ so as to measure such depressions.

The experiment is run as follows: an increasing amount of liquid $\Delta V_{w}$
is withdrawn from the ball through valve $\mathcal{V}_{s}$ (Fig.
\ref{fig:Principle}-b), via small volume intakes $\delta V_{i}$
($\Delta V_{w}=\sum\delta V_{i}$). After each step 
the ball is put in contact with the sole manometer through valve $\mathcal{V}_{m}$.
The displacement $h>0$ of the liquid in the manometer from the initial
equilibrium situation yields the pressure difference $P_{ext}-P_{int}=\rho gh$
across the ball membrane, where $\rho$ is the density of water. Because
of the fluid volume variations in the manometer, the inner volume
variation $\Delta V$ of the shell is slightly different from the
volume $\Delta V_{w}$ set through the syringe : 
\begin{equation}
\Delta V=\Delta V_{w}-\pi r^{2}h,\label{eq:vcorr}
\end{equation}
where $r$ is the internal radius of the (cylindrical) manometer tube.
Even though this correction is systematically taken into account,
the problem with large sections $S=\pi r^2$ would be that the volume withdrawn
in the syringe has to be much larger than the targeted $\Delta V$,
which may possibly make the system jump to another stability branch. This could impede full characterization of the branch of interest ; this
is discussed in detail in subsection \ref{sub:Equilibrium-and-manometer-1}.
On the other hand, the limitation when decreasing $S$ lies in
a possibly high equilibration time (see subsection \ref{sub:StabilsizationTimeExp}). These experimental
precautions being taken into account, for each ball the pressure difference
$\Delta P$ at mechanical equilibrium can be plotted
with respect to the relative volume variation $\frac{\Delta V}{V_{0}}$,
giving insights on the state equation $\wp(\frac{\Delta V}{V_0})$ that
is expected to depend on the relative thickness $d/R$ of the ball,
and on its material's properties ($Y_{3D}$, $\nu$).

The two-step procedure ensures to work at almost imposed volume and to discuss the time evolution of the system from a known state. Had the valves  $\mathcal{V}_{s}$ and  $\mathcal{V}_{m}$ always been kept open so as to measure simultaneously volumes and pressures, the interpretation of the dynamics towards equilibrium  would have been more tricky, since sucking out fluids in the manometer amounts to imposing pressure in the shell once the withdrawal step is stopped. The relative contribution of the volume withdrawal in the shell and in the manometer would depend on the whole set-up configuration, and in particular on the tubings resistance, as well as on the shell mechanical properties.

\section{Deflation of spherical shells}

Deflation essentially occurs within two regimes. In a first mode of
deformation, the ball roughly keeps its sphericity. Then a sudden
transition \cite{LandauBook,Knoche2011,Quilliet_2012,Hutchinson_2017,Church_1994,Hutchinson_2016}
transforms the sphere into an axisymmetric shape with a dimple (Fig.
\ref{fig:spi}). Further deflation makes the  dimple
size continuously increase (Fig. \ref{fig:photo-balles}-a) \cite{Knoche2011,Quilliet_2012,Hutchinson_2017}. Note that  quick deflation can lead to multi-dimple deformations, which were shown
to correspond to branches of higher energy \cite{Quemeneur2012}, but this was not observed
thanks to our small-stepped-deflation.

\subsection{Linear regime before buckling}
\label{sec:linreg}

The first regime corresponds to constraints with a spherical symmetry,
which results in a ``in-plane" compression of the shell (\textit{i.e}.
parallel to the free surfaces). For materials with nonzero Poisson's
ratio, this induces elongationnal shear in the thickness of the shell
but, in the surface model that is used to describe thin shells, spherical shrinking can be modelled by a uniform in-plane compression
of a spherical surface \cite{Quilliet_2012}. In a $\Delta P$ versus
$\frac{\Delta V}{V_{0}}$ diagram, quadratic compression energy corresponds
to a linear evolution \cite{Quilliet_2012,Marmottant_2011}: $\Delta P=\frac{4\chi_{2D}}{3R}\left(\frac{\Delta V}{V_{0}}\right)$,
where $\chi_{2D}$ is the surface compression modulus. For a thin
shell of an isotropic material, this 2D effective parameter can be
linked to the 3D properties of the shell through $\chi_{2D}=\frac{Y_{3D}d}{2\left(1-\nu\right)}$,
where $Y_{3D}$ is the Young modulus of the material, and $\nu$ its
Poisson's ratio ($\nu\lesssim0.5$ for most of the elastomeric materials,
these latter being exclusively used for our experiments because they
can undergo a 200\% elongation without plastic deformation or fracturation).
Hence:
\begin{equation}
\Delta P=\frac{2Y_{3D}}{3\left(1-\nu\right)}\times\frac{d}{R}\left(\frac{\Delta V}{V_{0}}\right).\label{eq:LinTheo}
\end{equation}

\begin{figure}[t]
\resizebox{\columnwidth}{!}{\includegraphics{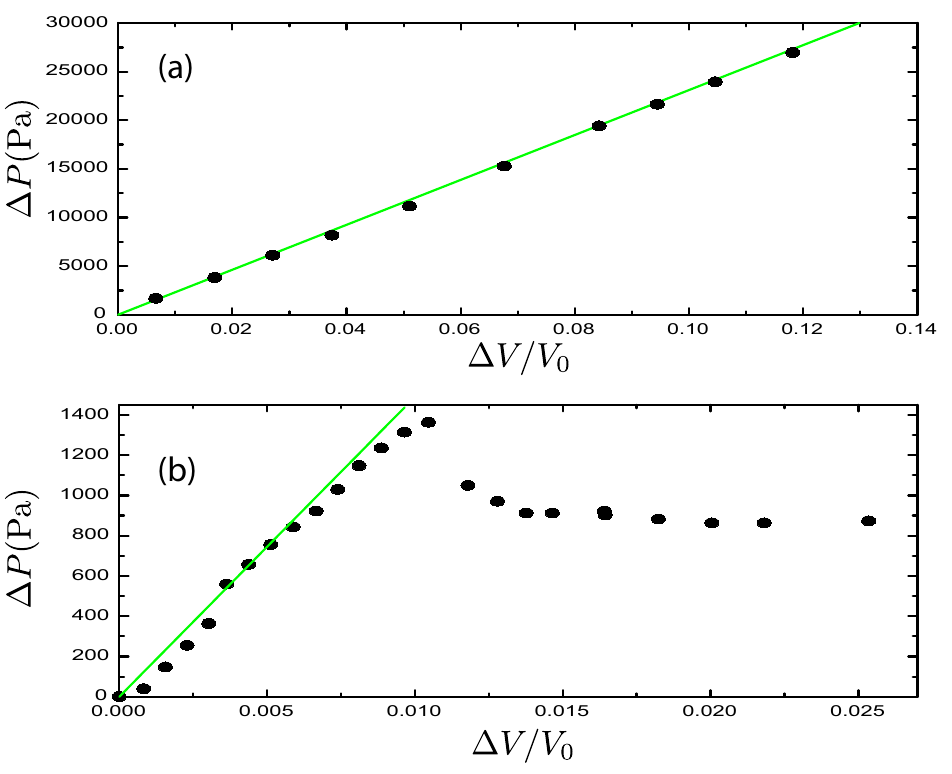}}
\caption{\label{fig:PvsVimm}Outside-inside pressure difference $\Delta P=P_{ext}-P_{int}$
for different elastic shells, versus the relative volume variation.
All measurements done at equilibrium for different water volume removals
are represented. Green line: linear fit before buckling. (a) $R=22.5$\,mm,
$d/R=0.222,$ $Y_{3D}=0.5$\,MPa (measured by traction experiments,
see subsection \ref{sec:Traction}). (b) $R=51.2$\,mm, $d/R=0.0293$,
$Y_{3D}=5.5$\,MPa. Buckling can be observed here for $\frac{\Delta V}{V_{0}}\,\apprge\,0.01$,
and plateauing (see subsection \ref{sub:Plateau}) for $\frac{\Delta V}{V_{0}}$
over about 0.015.}
\end{figure}

Experiments effectively show the expected linear behaviour,
as exemplified in Fig. \ref{fig:PvsVimm}. Values of the slope are
used to nondimensionalise the characteristic post-buckling pressures  in subsection \ref{sub:Plateau-values},
and are compared in section \ref{sec:Traction} to traction experiments
which provided independent measurements of $Y_{3D}$ and $\nu$.

This linear regime persists up to the point where
an instability causes a drastic change of shape ("buckling") toward
a configuration with a single axisymmetric dimple, together with a
drop of $\Delta P$. The critical pressure at which buckling takes place was predicted
from classical buckling theory \cite{Hutchinson 1967,Knoche2011}
to be:
\begin{equation}
\Delta P_{c}=\frac{2}{\sqrt{3\left(1-\nu^{2}\right)}}\times Y_{3D}\left(\frac{d}{R}\right)^{2}.\label{eq:Hutchinson}
\end{equation}
In experiments, buckling
often occurs before this threshold is reached, because of defects in the material
\cite{Vella2011,Reis2017}, possibly down to 20\% of the theoretical
predictions for a perfect shell \cite{Hutchinson_2016}.

According to numerical studies  \cite{Knoche2011,Quilliet_2012,Marmottant_2011}, proceeding with small deflation steps after this
buckling hardly changes the value of $\Delta P$, which roughly plateaus
during a substantial range of $\frac{\Delta V}{V_{0}}$.
Plateauing, which is exemplified in Fig. \ref{fig:PvsVimm}-b is specifically studied in the next section. For the
thinnest shells, further deflation steps lead to a second, softer
transition where radial folds progressively appear in the dimple (see Fig. \ref{fig:photo-balles}-b and
refs. \cite{Quilliet2008,Quilliet2008err,Quilliet_2012})
; this aspect is not  addressed in the present paper.

\begin{figure}
\resizebox{\columnwidth}{!}{\includegraphics{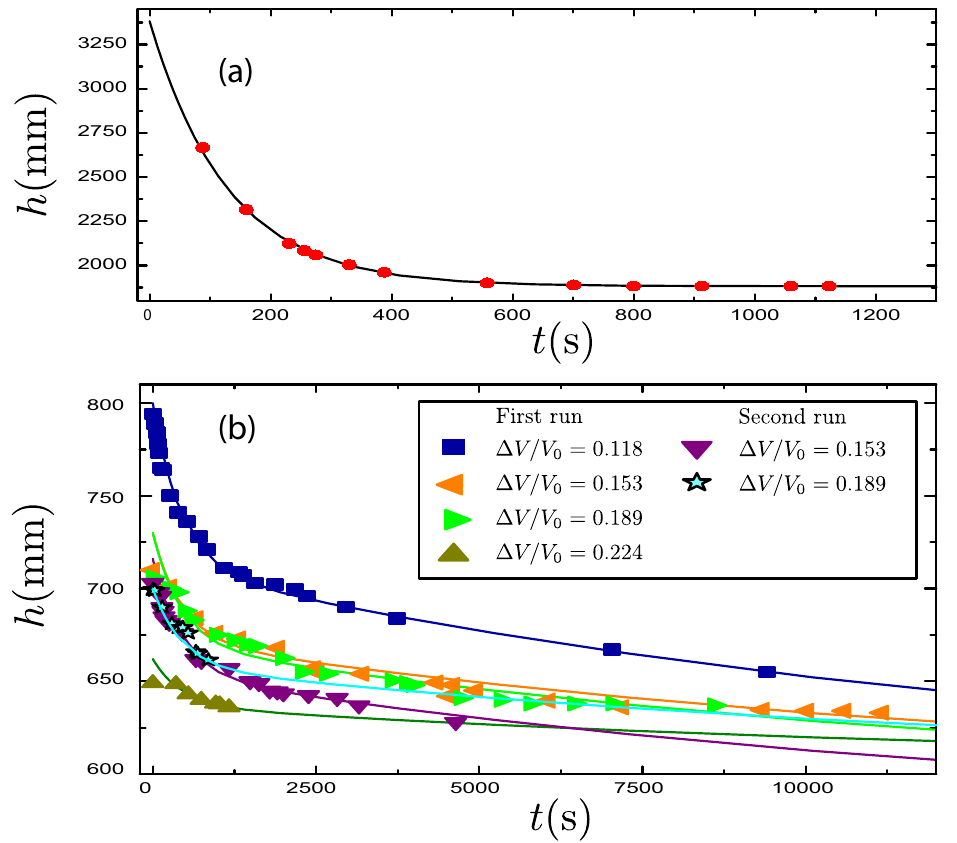}}
\caption{\label{fig:FitBiExp} "Slow devices": Time evolution of the slump $h$ of water level
in the manometer.
(a) Monoexponential relaxation with a characteristic time $\tau=129$\,s for a ball of radius $R=22.5$\,mm and relative thickness $d/R=0.222$; relative
volume variation after stabilization (\textit{i.e.} calculated using
$h_{asympt}$ in Eq. (\ref{eq:vcorr})) $\frac{\Delta V_{asympt}}{V_{0}}=0.211$.
(b) Biexponential relaxation for different values of the volume sucked
out from a ball of radius $R=58.7$\,mm and $d/R=0.097$, during 2 different
sequences of deflation. All theoretical curves correspond to a biexponential
fit \underline{with the same pair} of characteristic times ($\tau_{1}=440$\,s,
$\tau_{2}=15000$\,s) and \underline{the same proportion} $p=0.42$
of short-time exponential. }
\end{figure}

\subsection{Post-buckling plateau\label{sub:Plateau}}

\subsubsection{Stabilization time\label{sub:StabilsizationTimeExp}}

During the spherical mode of deflation, the
water level continuously falls (stabilizing within a few seconds)
every time a small amount of water is sucked out from the ball. After
buckling, it suddenly rises in the manometer. When deflation is performed
further on, several behaviours may take place:

- For most ball+manometer devices, the water level in the manometer
stabilizes within a few seconds at each post-buckling deflation step.
When recorded for a large range of relative volume variations, the
slump $h$ under the reference equilibrium level in the manometer
hardly varies with $\frac{\Delta V}{V_{0}}$ ("plateauing"). It
shows indeed a very weak minimum at some intermediate value (as examplified
in fig. \ref{fig:PvsVimm}-b for $\frac{\Delta V}{V_{0}}$ above 0.015). Then, for the experiments carried at sufficiently large deflation,
it re-increases which corresponds to an expected divergence when $\frac{\Delta V}{V_{0}}$
approaches 1 (ideally emptied ball)\cite{Knoche2011}. The minimum
value $h{}_{min}$ of $h$ when it plateaus allows to determine the
so-called "plateauing pressure" $\Delta P_{pl}=\rho gh_{min}$.
This quantity underwent a specific study in the numerical simulations
of ref. \cite{Quilliet_2012}, which will be revisited hereafter.

- Nevertheless, for some ball+manometer devices, at all deflation
steps $h$ systematically shows a steep increase (\textit{i.e}. the
water level is suddenly sucked down for a few seconds) every time the ball is reconnected
to the manometer after the sucking out of $\delta V_{i}$; then it
decreases during minutes or more before stabilization, down to a new
equilibrium value $h_{asympt}$. In the following, these experimental configurations are named "slow devices". For most shells where post-buckling
equilibrium is not immediately realized, it would have been too long
to wait for $h$ reaching the $h_{asympt}$ value for each relative
volume variation $\frac{\Delta V}{V_{0}}$ explored. Fortunately,
we found out that the decrease of $h(t)$ was exponential
for a few cases (fig. \ref{fig:FitBiExp}-a), and that in the other
cases it could be fitted using a biexponential of general formula:
\begin{equation}
h\left(t\right)=h_{asympt}+\left(h_{init}-h_{asympt}\right)\left[p\,e^{-t/\tau_{1}}+\left(1-p\right)e^{-t/\tau_{2}}\right],\label{eq:BiExpRelation}
\end{equation}

where $\tau_{1}$ and $\tau_{2}$ are respectively the short and long
characteristic times, and $p$ the proportion of short-time exponential
in the modelled signal (see Fig. \ref{fig:FitBiExp}-b).

Mechanical equilibrium is realized only when the water level in the
manometer reaches its asymptotic value $h_{asympt}$. The $\left(\frac{\Delta V}{V_{0}},\Delta P=\rho gh_{asympt}\right)$
experimental graph shows  plateauing as for balls
without time delay. Results are presented and discussed in subsection
\ref{sub:Plateau-values}.

\begin{figure}
\resizebox{\columnwidth}{!}{\includegraphics{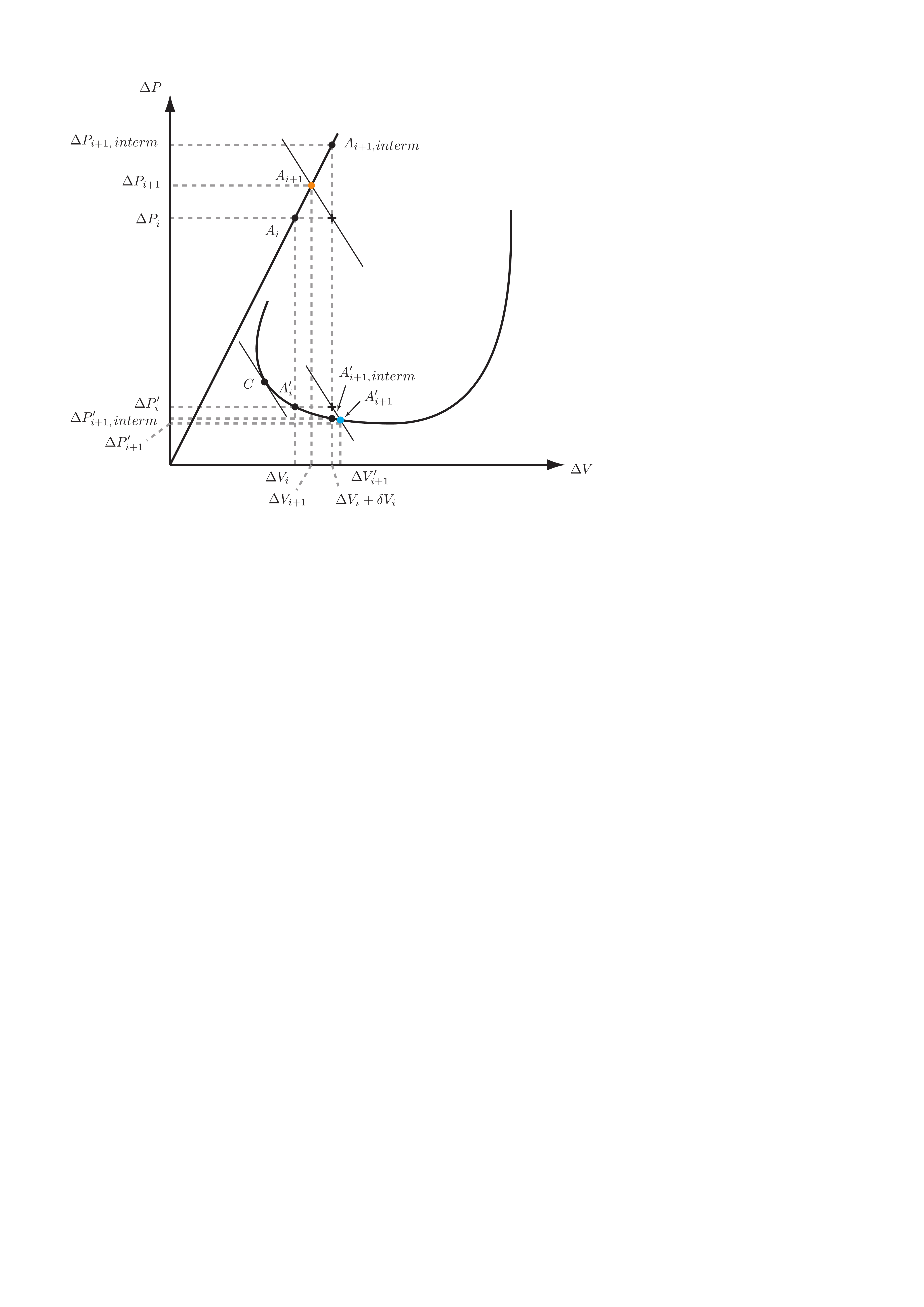}}

\caption{\label{fig:TransientStab}Continuous black lines: typical state function (see \cite{Knoche2011,Quilliet_2012}),
that can locally be denotated as $\wp(\Delta V)$. Upper branch: after
removal of a volume of water $\delta V_{i}$ (with the syringe through
valve $\mathcal{V}_{s}$) from the equilibrium state $A_{i}\left(\Delta V_{i},\Delta P_{i}\right)$,
the shell finds a non-observable intermediate equilibrium state $A_{i+1,interm}\left(\Delta V_{i}+\delta V_{i},\Delta P_{i+1,interm}\right)$.
Opening valve $\mathcal{V}_{m}$ for the measurement of the inner
pressure leads to a new (observable) equilibrium state $A_{i+1}\left(\Delta V_{i+1},\Delta P_{i+1}\right)$,
in orange, at the intersection of $\wp(\Delta V)$ and the straight line
of slope $\left(-\frac{\rho g}{\pi r^{2}}\right)$ (cf Eq. (\ref{eq:DteFonctionnt}))
that passes through the point $\left(\Delta V_{i}+\delta V_{i},\Delta P_{i}\right)$).
Since $\frac{d\wp}{d(\Delta V)}>0$, one gets $\Delta V_{i+1}<\Delta V_{i}+\delta V_{i}$.
Lower branch: same construction (with primed symbols) for intermediate and final state (in blue). As $\frac{d\wp}{d(\Delta V)}<0$ there,
the inequality reverts: $\Delta V_{i+1}>\Delta V_{i}+\delta V_{i}$.
This construction shows that on the lower stability branch, states
corresponding to a $\Delta P$ higher than in state C, where the slope
equals to $\left(-\frac{\rho g}{\pi r^{2}}\right)$, cannot be explored. }
\end{figure}

\subsubsection{Equilibrium and manometer\label{sub:Equilibrium-and-manometer-1}}

The equilibrium configurations, and the route toward them, are obtained while the shell is in contact with the manometer. In the two following subsections, we establish how this coupling influences the way the state diagram is explored and how the dynamical features intrinsic to the shell can be extracted.

After closing of valve $\mathcal{V}_{s}$
and opening of valve $\mathcal{V}_{m}$ (see Fig. \ref{fig:Principle}),
pressure adaptation between the ball and the manometer occurs through
water exchange, which in turn modifies (i) the pressure exerted by
the water column in the ball (ii) the volume of the ball, hence the
pressure exerted by the shell. The final state emerges from this feedback. Two characteristic situations are displayed in Fig. \ref{fig:TransientStab}.
After a volume $\delta V_{i}$ has been sucked
out from a ball at equilibrium with state $\left(\Delta V_{i},\Delta P_{i}\right)$, the ball finds itself in a state $\left(\Delta V_{i}+\delta V_{i},\Delta P_{i+1,interm}\right)$, which we assume here to be an equilibrium state.
Nevertheless, features of this new state are not known by the
experimentalist, who has to open valve $\mathcal{V}_{m}$ in order
to measure the pressure. Once the ball and manometer are in contact, the pressure
difference $\Delta P_{i+1,interm}$ between both extremities of
the manometer is not \textit{a priori} equilibrated by the water withdrawal
$h_{i}$ (that previously equilibrated $\Delta P_{i}$). This leads to a flow in the manometer until the outside-inside pressure difference
$\Delta P=P_{ext}-P_{int}$  is equilibrated by the hydrostatic pressure associated with withdrawal $h$: $\Delta P_{i}-\Delta P=\rho g\left(h_{i}-h\right)$. On an other
hand, conservation of water volume implies that $\Delta V-\Delta V_{i}=\delta V_i-\pi r^{2}\left(h-h_{i}\right)$;
hence:
\begin{equation}
\Delta P=\Delta P_{i}+\frac{\rho g}{\pi r^{2}}\left(\Delta V_{i}+\delta V_{i}-\Delta V\right).\label{eq:DteFonctionnt}
\end{equation}

\begin{figure}
\resizebox{\columnwidth}{!}{\includegraphics{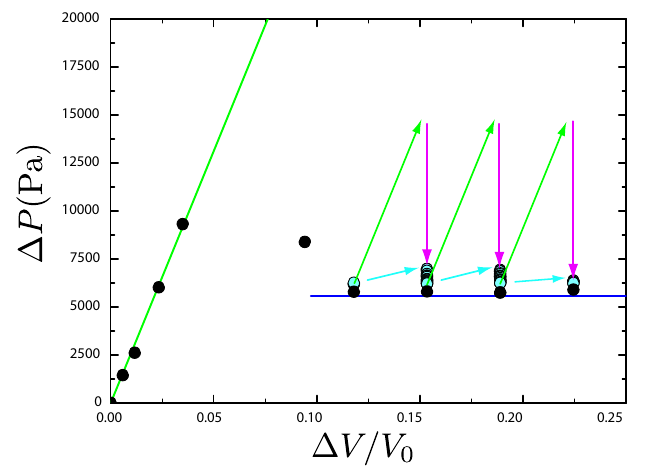}}
\caption{\label{fig:PvsVdelay}Black dots: inside-outside pressure difference
$\Delta P$ for the first deflation of the ball of Fig. \ref{fig:FitBiExp}-b, versus its relative volume variation$\frac{\Delta V}{V_{0}}$. For points after the buckling ($\frac{\Delta V}{V_{0}}>0.05$), the
pressure is calculated from the asymptotic value of $h\left(t\right)$
obtained through the biexponential fit such as in Fig. \ref{fig:FitBiExp}-b.
Blue disks represent $\rho gh$ for measurements of $h\left(t\right)$
performed before stabilization of the water level in the manometer;
this quantity, calculated out of equilibrium, does not correspond
to the pressure difference through the ball, but its representation
provides an estimation of the error performed if equilibrium is not
attained. It is to be noted that the variation of water height in
the manometer during the stabilization does not affect $\frac{\Delta V}{V_{0}}$
by a perceptible amount. In green and blue, respectively: the reconstructed
linear part of the state equation (spherical deformation, Eq. (\ref{eq:LinTheo}))
up to the critical pressure difference (Eq. (\ref{eq:Hutchinson})),
and the plateauing value of the post-buckling regime as proposed by
heuristic Eq. (\ref{eq:Magic Formula}). Arrows are associated with the
discussion at the end of subsection \ref{sub:Equilibrium-and-manometer-1}.}
\end{figure}

In a $\Delta P-\Delta V$ diagram, this is the equation of the straight
line ("operating curve") of slope $\left(-\frac{\rho g}{\pi r^{2}}\right)$
that passes through the point $\left(\Delta V_{i}+\delta V_{i},\Delta P_{i}\right)$
(Fig. \ref{fig:TransientStab}). The measured equilibrium state $\left(\Delta V_{i+1},\Delta P_{i+1}\right)$
is then found by following the state curve $\wp(\Delta V)$
from the intermediate equilibrium state (with valve $\mathcal{V}_{m}$
 closed) $(\Delta V_{i}+\delta V_{i},\Delta P_{i+1,interm})$
up to its intersection with the straight line of equation (\ref{eq:DteFonctionnt}).
Of course, if several branches of the state function are intersected,
the final state is expected to lie on the same branch as that reached
by the intermediate state (see Fig. \ref{fig:TransientStab}). Two limit cases
for the operating curve are horizontality, which marks deformations
at imposed pressure difference, and verticality (imposed volume).
Comparing the slopes of the linear part of the $\Delta P-\Delta V$
diagram and of the operating curve provides a threshold value $r_{c}=\left(\frac{\rho g R^{4}}{dY_{3D}}\right)^{1/2}$
for the inner radius of the manometer, so that $r\ll r_{c}$ corresponds
to deflation at imposed volume, and $r\gg r_{c}$ to deflation at
constant pressure. For our experimental conditions, $r_{c}\approx1\,$mm:
experiments are done in an intermediate regime where, in particular,
the jump between the two states before and after buckling has a negative
slope whose absolute value is comparable to the slope of the isotropic part of the deflation (see Figs. \ref{fig:PvsVimm}-b and  \ref{fig:PvsVdelay}).

The interplay between the shell and the manometer also sets a limitation
for the determination of the state function $\wp(\frac{\Delta V}{V_0})$: only the part of the lower branch corresponding
to $\Delta V>\Delta V_{C}$, where $C$ is the point where the tangent
has a slope $\left(-\frac{\rho g}{\pi r^{2}}\right)$ (Fig. \ref{fig:TransientStab}),
can be explored. Also the access to the extremity of the linear part
depends on $r$. Finally, a small internal radius $r$ of the
manometer allows to explore a bigger part of both the lower and upper
branches. The counterpart lies in the dynamics toward equilibrium,
which is discussed in the following subsection.

The situation is indeed more complex for some "slow devices" (ball+manometer)
where, in the post-buckling state, the equilibrium takes more than a few seconds
to stabilize after the opening of valve $\mathcal{V}_{m}$. In that case, water outtake generates a
steep withdrawal of the water level in the manometer, followed by a
slower increase toward a limit value (\textit{via} an exponential or
bi-exponential relaxation versus time, as exposed in subsection \ref{sub:StabilsizationTimeExp}).
We observed experimentally that the slope of the steep withdrawal (light
blue in fig. \ref{fig:PvsVdelay}) never overtakes the slope of the
linear part (which corresponds to pure constriction of the shell). We then assume that the sucking
out of $\delta V_{i}$ first generates a (rapid) uniform constriction
of the surface (on the figure: green arrows with the same
inclination than the linear part of $\wp(\frac{\Delta V}{V_0})$), which
has enough time to partly relax via a rolling of the rim that encircles
the depression (pink arrows) before the ball
is reconnected to the manometer. The relaxation of $h$ observed
afterwards, then, corresponds to the end of the rim rolling toward
the $\left(\Delta V_{i+1},\Delta P_{i+1}\right)$ equilibrium configuration,
possibly slowed down further by other phenomena discussed in the
following subsection.

A quantitative model for identifying the origin of the characteristic
time(s) that are observed after connection to the manometer is proposed
in the next subsection.
\begin{figure}
\resizebox{\columnwidth}{!}{\includegraphics[width=7cm]{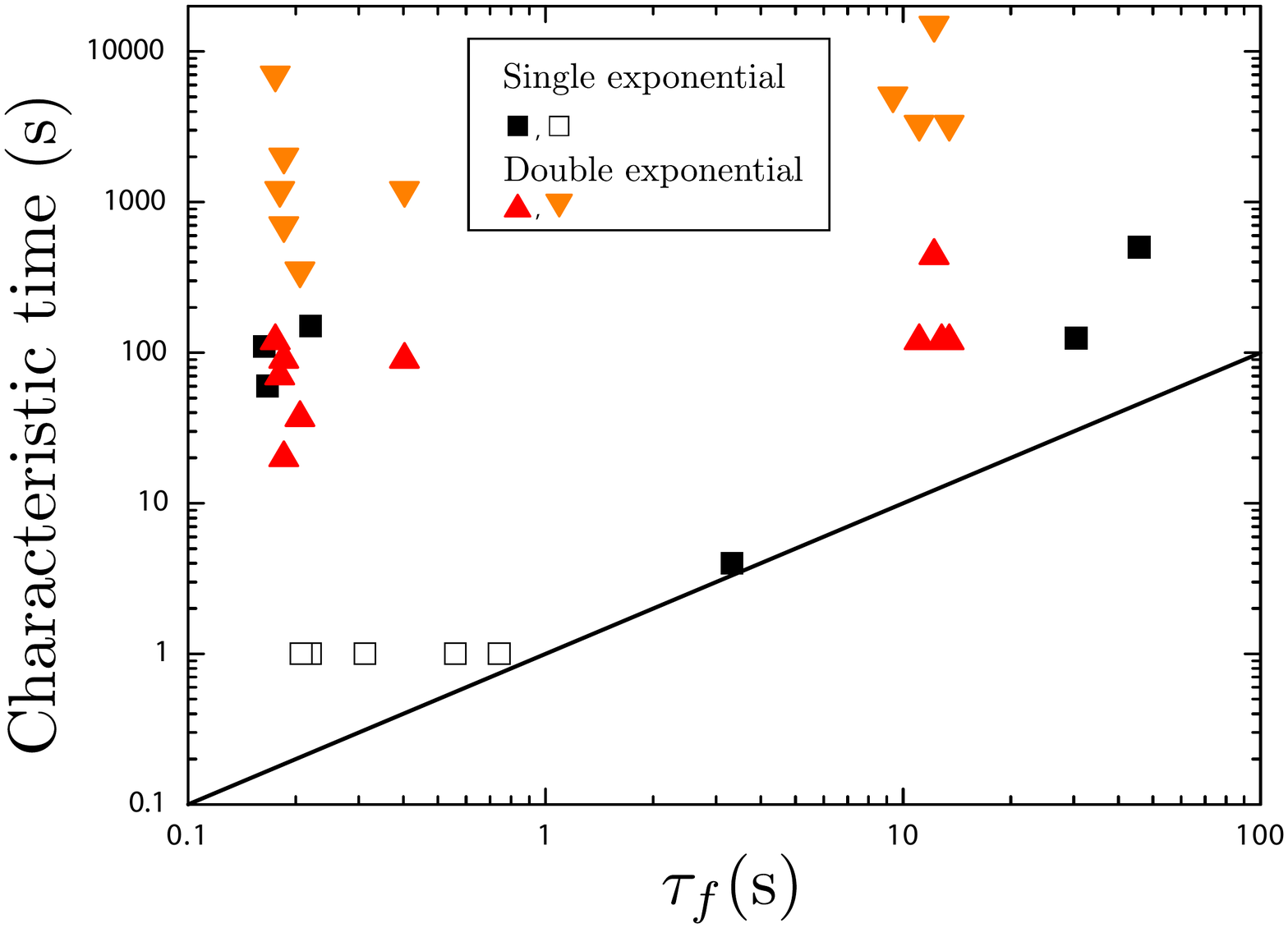}}

\caption{\label{fig:TempsEquilibrage}Characteristic equilibration times obtained
from mono- or biexponential fit of experimental $h\left(t\right)$
curves, with respect to $\tau_{f}$, the time associated to viscous
dissipation in the manometer, calculated using Eq. (\ref{eq:TpsDissViscLiq}).
Open squares: upper bound for the characteristic time of experiments
leading to ``immediate" equilibrium. Filled squares: unique characteristic time for
shells showing a monoexponential decay during deflation. Upward red
(resp. downward orange) triangles: short (resp. long) characteristic
time for shells showing biexponential decay during deflation. The
line indicates where the experimental times and $\tau_{f}$ (determined
by Eq. (\ref{eq:TpsDissViscLiq})) would meet.}
\end{figure}

\subsubsection{Relaxation towards equilibrium\label{sub:Model-for-the}}

As shown in figure \ref{fig:TempsEquilibrage}, the characteristic time is around $2-500\,s$
for an exponential decay while when a biexponential fit is necessary,
it unveils a longer characteristic time of $\approx300-20000\,s$.
Of course, for experiments where the water level stabilized "immediately",
we only have an upper bound for the characteristic time(s), which is
the few seconds that are necessary to operate the valves before measuring
$h$.

When valve $\mathcal{V}_{m}$ is turned open after a deflation step,
the water level in the manometer has to move in order to adapt to
the new pressure. Assuming
a Stokes incompressible flow in the vertical tube due to pressure difference
$\Delta P$ between both extremities, this writes:

\begin{equation}
8\eta\left(L-h(t)\right)\frac{dh(t)}{dt}- r^{2}\Delta P(t)+r^{2}\rho gh(t)=0,\label{eq:dyn-hp}
\end{equation}

where $\eta$ is the viscosity of the water, $L$ the total length
of the manometer, \textit{i.e.} from the
ball entry to the position of the meniscus at initial state. We neglected
the section variations at the level of the valves and connections,
and in the following we will replace $L-h$ by $L$ because $h\ll L$.
Rewriting Eq. (\ref{eq:dyn-hp}) then leads to:

\begin{equation}
\tau_{f}\frac{dh(t)}{dt}+h(t)=\frac{\Delta P(t)}{\rho g},\label{eq:dyn-h}
\end{equation}

where $\tau_{f}$ is a caracteristic time for the decay of the water
level toward its equilibrium value, and depends on experimental parameters
through:
\begin{equation}
\tau_{f}=8\eta L/(\rho gr^{2}).\label{eq:TpsDissViscLiq}
\end{equation}

However, this fluid viscous dissipation is not the only possible contribution
to the water level dynamics. As exposed in the end of the previous
subsection, internal frictions in the material that forms the shell
may be of importance. Our assessment is that, because of dissipation
in the shell's material, the pressure difference between both sides
of the shell may evolve with a characteristic time $\tau_s$ toward the equilibrium
situation where $\Delta P=\wp(\Delta V)$: 
\begin{equation}
\tau_{s}\frac{\Delta P(t)}{dt}+\Delta P(t)=\wp(\Delta V(t)).\label{eq:dptdvt}
\end{equation}

Here, we assume that $\tau_{s}$ is independent from the shape along
the equilibration process in the manometer, which is reasonable as
soon as small volume variations are imposed at each measurement step.

When opening valve $\mathcal{V}_{m}$ in order to measure the pressure,
the system evolves from the intermediate state $(\Delta V_{i},$ $\Delta P_{i+1,interm})$
to the state $\left(\Delta V_{i+1},\Delta P_{i+1}\right)$; equations
(\ref{eq:dptdvt}) and (\ref{eq:dyn-h}) together with the relationship
$\Delta V=\Delta V_{i}+\delta V_{i}+\pi r^{2}(h_{i}-h)$ eventually
lead to the evolution equation for $h$: 
\begin{multline}
\tau_{f}\tau_{s}\frac{d^{2}h(t)}{dt^{2}}+(\tau_{f}+\tau_{s})\frac{dh(t)}{dt}+h(t)\\
=\frac{\wp(\Delta V_{i}+\delta V_{i}+\pi r^{2}(h_{i}-h(t)))}{\rho g}.\label{eq:h}
\end{multline}

Before going further in the study of the dynamics towards measurable
equilibrium states, let us focus on the latter, which we denote with
stars. These states are characterized by hydrostatic relationship $\wp(\Delta V^{*})=\rho gh^{*}$,
with: 

\begin{equation}
\Delta V^{*}=\Delta V_{i}+\delta V_{i}+\pi r^{2}(h_{i}-h^{*}).\label{es:stat}
\end{equation}

Because we explore the diagram step-by-step, the system is never
far from its fixed point (except at the  moment of  exact buckling,
that we do not consider here), so that we can expand the second term
of equation (\ref{eq:h}) around it : $\wp(\Delta V)=\wp(\Delta V^{*})+\frac{d\wp}{d\Delta V}(\Delta V^{*})\times(\Delta V-\Delta V^{*})$,
and eventually:

\begin{multline}
\tau_{f}\tau_{s}\frac{d^{2}h(t)}{dt^{2}}+(\tau_{f}+\tau_{s})\frac{dh(t)}{dt}\\
+\big[1+\frac{\pi r^{2}}{\rho g}\times\frac{d\wp}{d\Delta V}(\Delta V^{*})\big](h(t)-h^{*})\\
=\frac{\wp(\Delta V^{*})}{\rho g}-h^{*}.\label{eq:hfinal}
\end{multline}

Initial conditions at $t=0$ are $h=h_{i}$, and from Eq (\ref{eq:dyn-h}), $\tau_{s}\frac{dh}{dt}=\frac{\Delta P}{\rho g}-h_{i}=\frac{\Delta P_{i+1,interm}}{\rho g}-h_i$, which depends on the moment at which the manometer was put in contact with the shell.

\begin{figure}[t]
\resizebox{\columnwidth}{!}{\includegraphics{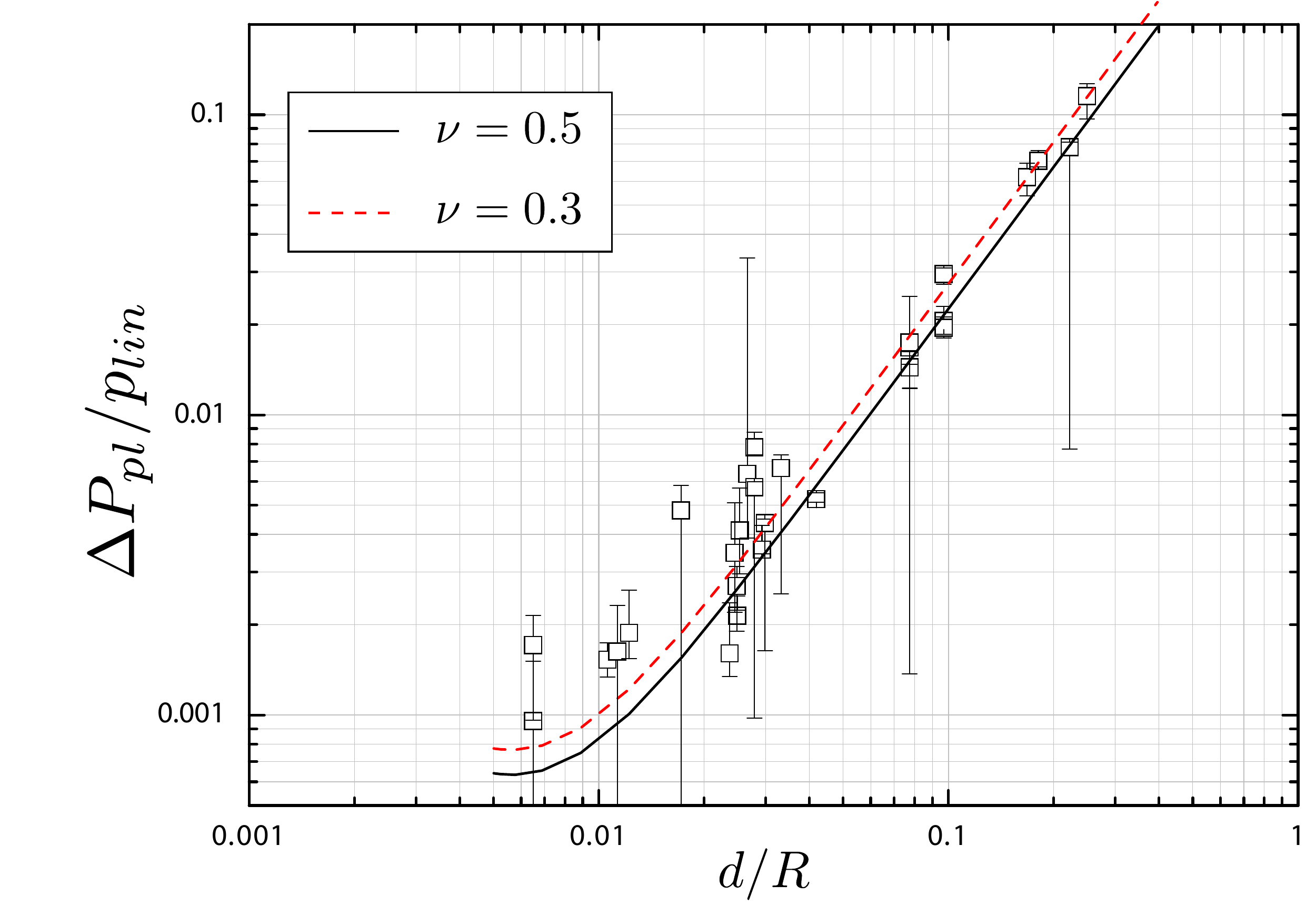}}
\caption{\label{fig:Resum1}Squares: values of $\Delta P_{pl}/p_{lin}$, obtained
from deflation curves similar to Fig. \ref{fig:PvsVimm}-b or Fig.~\ref{fig:PvsVdelay},
where $\Delta P_{pl}$ is the minimum value of the post-buckling regime,
normalized by the slope $p_{lin}$ of the pre-buckling linear part.
Lines: theoretical values determined using Eq. (\ref{eq:magic2}).}
\end{figure}

\begin{figure*}[t]
\begin{center}
\resizebox{1.9\columnwidth}{!}{\includegraphics{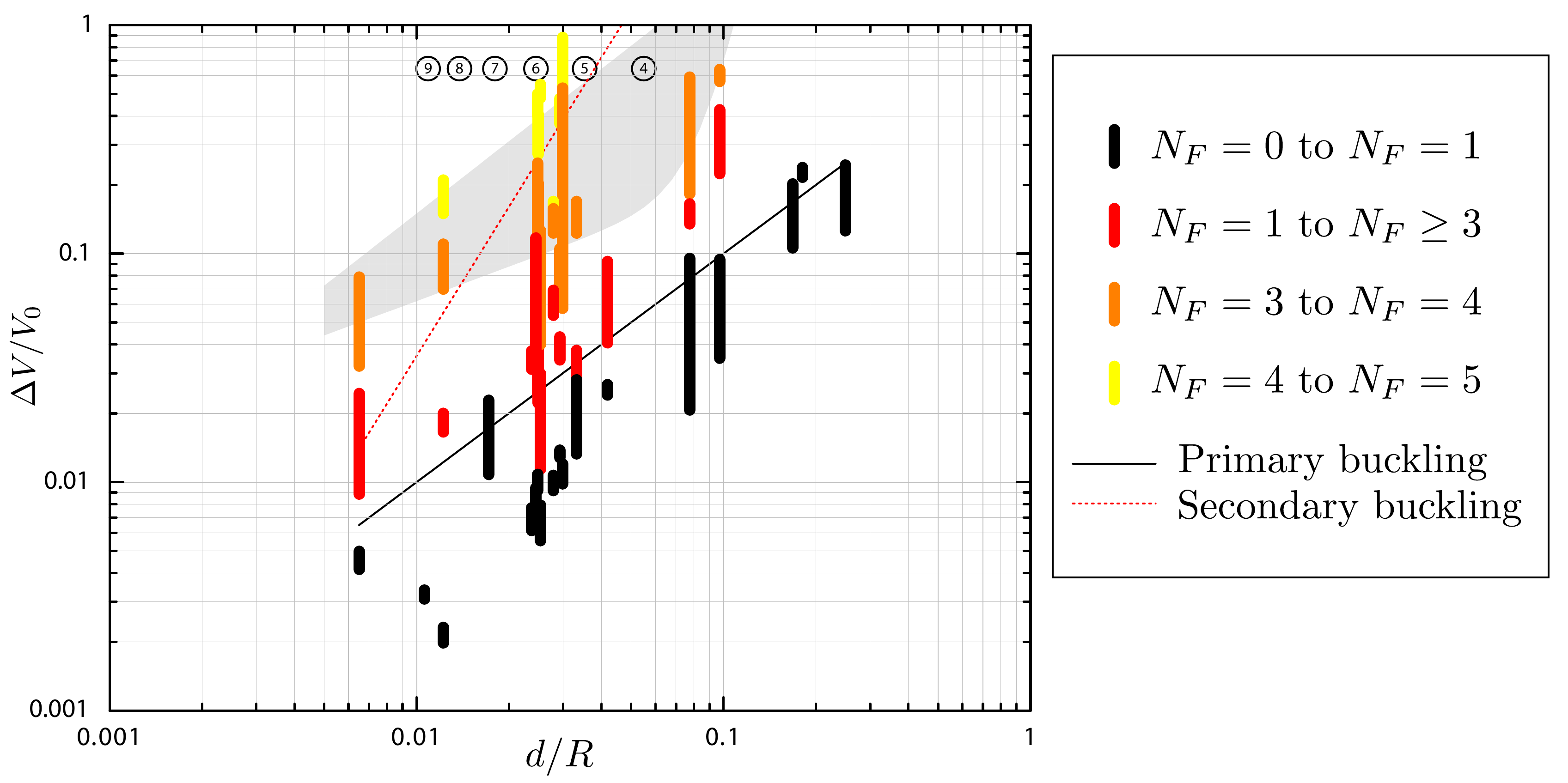}}
\caption{\label{fig:folds}
Shape diagram during deflation in the $(d/R,\Delta V/V_0)$ space. Thick lines indicate the  $\Delta V/V_0$ value range for which the considered transition was experimentally observed during deflation. Black: first buckling transition. Red: second buckling transition (from $N_F=1$ to $N_F\ge 3$). Orange: transition to $N_F=4$. Yellow: transition to $N_F=5$. The grey area is an indicator  of the domain of existence of the $N_F=4$ configuration. Black thin line indicates the theoretical boundary for the primary transition for $\nu=0.5$ (see. Eq. 8 in Ref. \cite{Quilliet_2012}). Red dashed line indicate the secondary transition as obtained from Eq. 11 in \cite{Quilliet_2012} with $\nu=0.5$. Theoretical boundaries depend only weakly on $\nu$.
In the black circles are indicated the expected values for $N_F$ at the end of the plateau ($\Delta V/V_0\in[0.53;0.76]$), which are displayed in Fig. \ref{fig:NF}.}
\end{center}
\end{figure*}

\begin{figure}[t]
\resizebox{\columnwidth}{!}{\includegraphics{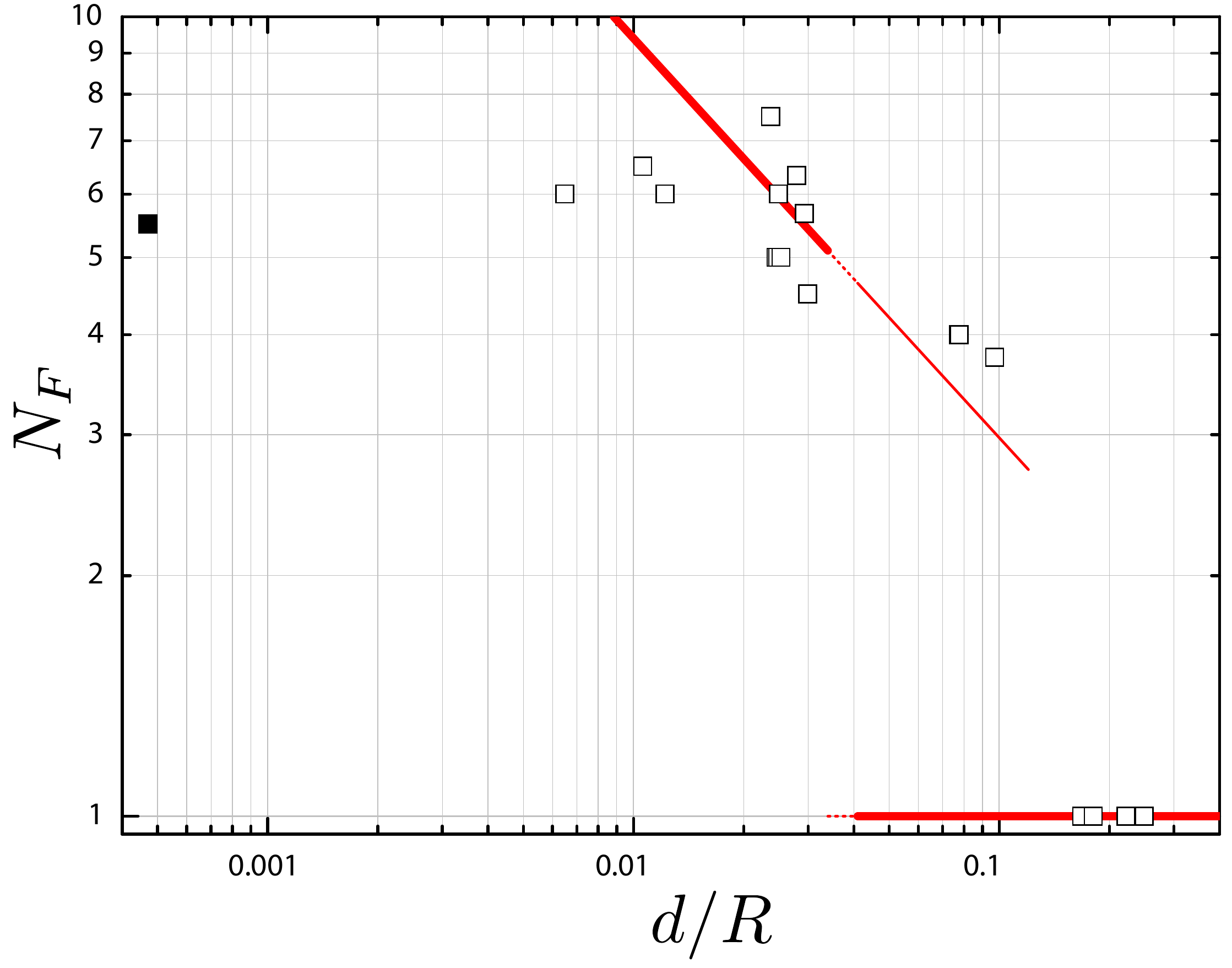}}
\caption{\label{fig:NF}
Number $N_F$ of folds, averaged for observations between $\Delta V/V_0=0.53$ and 0.76 (open squares). Full square is the experimental point of Ref. \cite{Carlson_1968}.  $N_F=1$ stands for "axisymmetric dimple". 
The thick line indicates the  expected $N_F= 0.94\times (d/R)^{-1/2}$, obtained from the analysis of simulations in the same deflation range \cite{Quilliet_2012}. The dashed line  corresponds to situations where the secondary buckling takes place in the $[0.53,0.76]$ range for $\Delta V/V_0$. The thin line corresponds to an extrapolation of the heuristic law  in the $d/R$ range where simulations did not predict the secondary buckling.}
\end{figure}

One can easily show that the characteristic equation associated with the left part of Eq. \ref{eq:hfinal}
has two roots with negative real parts if $\left[\frac{d\wp}{d\Delta V}(\Delta V^{*})\right]>\left[-\frac{\rho g}{\pi r^{2}}\right]$. If this is not the case, the fixed point is not a stable point and
cannot be reached, as already discussed in the geometrical construction of Fig. \ref{fig:TransientStab}. This implies we cannot explore parts of the $\wp(\Delta V)$
state function where the slope is too strongly negative. Those are
scarce in the diagram \cite{Knoche2011}, which justifies the choice
of a U-shape manometer with water below the air at the level of the
interface.

The strongest slopes are met in the isotropic phase. In that case,
$\frac{d\wp}{d\Delta V}(\Delta V^{*})\sim Y_{3D}\times\frac{d}{R}\times\frac{1}{V_{0}}$.
Considering $r=0.5$ mm, $Y_{3D}=7$ MPa, the highest value 0.25 for
$d/R$ and the lowest value $40$ mm for the shell radius, we find
that $\frac{\pi r^{2}}{\rho g}\times\frac{d\wp}{d\Delta V}(\Delta V^{*})$
never exceeds 0.5, so this term can be safely ignored in Eq. (\ref{eq:hfinal}) when one studies post-buckling states.

For the isotropic phase as for the plateau, the solution of Eq. \ref{eq:hfinal}
is therefore a biexponential function with characteristic times $\tau_{f}$
and $\tau_{s}$.

The theoretical $\tau_{f}$, calculated using Eq. (\ref{eq:TpsDissViscLiq}),
was compared (Fig. \ref{fig:TempsEquilibrage}) to the characteristic
time(s) experimentally obtained, to which we incorporated data from
the "instantaneous" experiments by estimating the upper bond for
the characteristic time as $1\,$s. We
observe that apart from one case, the characteristic times are
much higher than the viscous time $\tau_{f}$. This suggests that
the times experimentally determined are intrinsic to the shells
themselves, which are made of commercial polymers. Eq. \ref{eq:dptdvt} needs to be refined to account for this more complex relaxation scenario, which depends strongly on the ball under consideration as 0, 1 or 2 characteristic times larger than a few seconds can emerge.
One may wonder why the viscous fluid characteristic time was not observed
in more cases:  the sampling 
was adapted to the slow relaxation dynamics of the shells, preventing data collection at times necessary to detect exponential
contribution(s) with a characteristic time of a few seconds.

Finally, the choice of an intermediate section for the manometer enables us to obtain fluid dissipation times well-separated from that associated with the dissipation in the shell material, without hindering our ability to explore the state diagram by the use of too large sections.

\subsubsection{Plateau values\label{sub:Plateau-values}}

We denominate by $\Delta P_{pl}$ ("plateau value") the minimum value of the
outside-inside pressure difference $\Delta P$, in the very flattened
U-shaped part of the curve after buckling.

This quantity was previously studied through numerical simulations
in Ref. \cite{Quilliet_2012}, and a heuristic dependance had been
found between $\Delta P_{pl}$ and $\frac{\Delta V}{V_{0}}$. For
the present paper, we extended the simulations range and we use a different
formula to fit the simulations for the whole range of experimental
$\frac{d}{R}$, \textit{i.e.} from $5.10^{-3}$ to 0.3:

\begin{equation}
\Delta P_{pl}=\frac{Y_{3D}}{\left(1-\nu^{2}\right)^{0.75}}\times\left(2.34\,10^{-6}+0.9\left(d/R\right)^{2.57}\right)\label{eq:Magic Formula}
\end{equation}

In order to check  the consistency of the deflation experiments
with the theory, we determined for each ball the slope $p_{lin}$
of the linear part. Theoretically, $p_{lin}=\frac{2Y_{3D}}{3\left(1-\nu\right)}\times\frac{d}{R}$
(from Eq. (\ref{eq:LinTheo})). We then focussed on the nondimensionalized
value $\frac{\Delta P_{pl}}{p_{lin}}$ (which avoids concerns about
an independent determination of $Y_{3D}$) with respect to $\frac{d}{R}$,
as displayed in Fig. \ref{fig:Resum1}. It shows
that these experimental points are consistent with the theoretical
curve obtained from Eq. (\ref{eq:Magic Formula}) and the expression
of $p_{lin}$:
\begin{equation}
\frac{\Delta P_{pl}}{p_{lin}}=\frac{\left(1-\nu\right)^{0.25}}{\left(1+\nu\right)^{0.75}}\left[3.51\,10^{-6}+1.35\left(\frac{d}{R}\right)^{2.57}\right]\left(\frac{d}{R}\right)^{-1}\label{eq:magic2}
\end{equation}

This result is new and of practical interest, since equation (\ref{eq:Magic Formula})
had been established for thin shells. The experiments presented
here show its validity for shells with relative thickness up to $\frac{d}{R}\approx0.3$.

\subsubsection{Towards folding\label{sub:folds}}

As the ball deflates along the postbuckling plateau, folds appear  in the depression for the thinnest of the shells, as in Fig. \ref{fig:photo-balles}-b. This secondary buckling transition is documented in literature for thin shells, both experimentally \cite{Carlson_1968} for very thin shells and theoretically \cite{Knoche2014,Hutchinson_2017}, but the only results for what concerns shells of medium thickness ($d/R>0.02$) were obtained numerically \cite{Quilliet2008,Quilliet_2012}.
Experimental domains of existence of emblematic non axisymmetric conformation are represented in Fig. \ref{fig:folds} in the $(d/R,\Delta V/V_0)$  space. They show some discrepancies with the boundaries obtained from simulations  \cite{Quilliet2008,Quilliet_2012}.

Primary buckling occurred for a volume loss much lower than that predicted in simulations ;  as in Sec. \ref{sec:linreg}, defects are expected to be the cause of this discrepancy.

The secondary buckling towards non-axisymmetric shapes also occurred for values of the relative deflation significantly lower than in simulations. Such shapes present radial folds, the number of which is denominated by $N_F$. In our experiments, the transition out from an axisymmetric shape could happen by way of an elongation of the dimple (the shape is then characterized by $N_F=2$, as in Ref. \cite{Carlson_1968}) and could be continued by the development of three fold shape ($N_F=3$). Both types of shapes were not obtained in the simulations of Ref. \cite{Quilliet_2012}. In these simulations, $N_F=4$ was seldom observed while the $N_F=4$ zone shows a great extent in the experimental  diagram of Fig. \ref{fig:folds}. Finally, the experimental domain of transition from $N_F=3$ to $N_F=4$ is crossed by the heuristic transition line found in Ref. \cite{Quilliet_2012}Ê for the secondary buckling, that is characterized by a $N_F=1$ to $N_F\ge 4$ direct transition. It may indicate that, for some numerical reason, the energy minima corresponding to low numbers of folds were not found by the solver in the simulations, that were then stuck to axisymmetric shapes. Note that the secondary buckling transition line found by Knoche and Kierfeld in Ref. \cite{Knoche2014} is close to that proposed in Ref. \cite{Quilliet_2012}, that serves here as a reference for this discussion.

Notwithstanding this discrepancy in the boundaries of the axisymmetric zone, we aim here at checking the heuristic dependence with $d/R$ of the number of folds $N_F$ reached at the end of the plateau,  proposed in Ref. \cite{Quilliet_2012}.

For the thinnest of the shells, the number of folds clearly departs from this heuristic law, as shown in Fig. \ref{fig:NF}. This discrepancy may be due to the intrinsic limitations of an elastic model, failing to describe microscopic phenomena at stake at the apex of the s-cones in thin shells, where sharp creases are likely to host plastic deformation \cite{Nasto2013}.
Interestingly, for thick enough shells ($d/R > 0.01$) less prone to extreme deformations, the number of folds roughly follows the proposed law in $(d/R)^{-1/2}$, thus confirming the relevance of $\sqrt{dR}$ as the key length for the elastic deformations of shells \cite{Quilliet_2012}.

\section{Comparison with traction experiments\label{sec:Traction}}

\begin{figure}
\resizebox{\columnwidth}{!}{\includegraphics{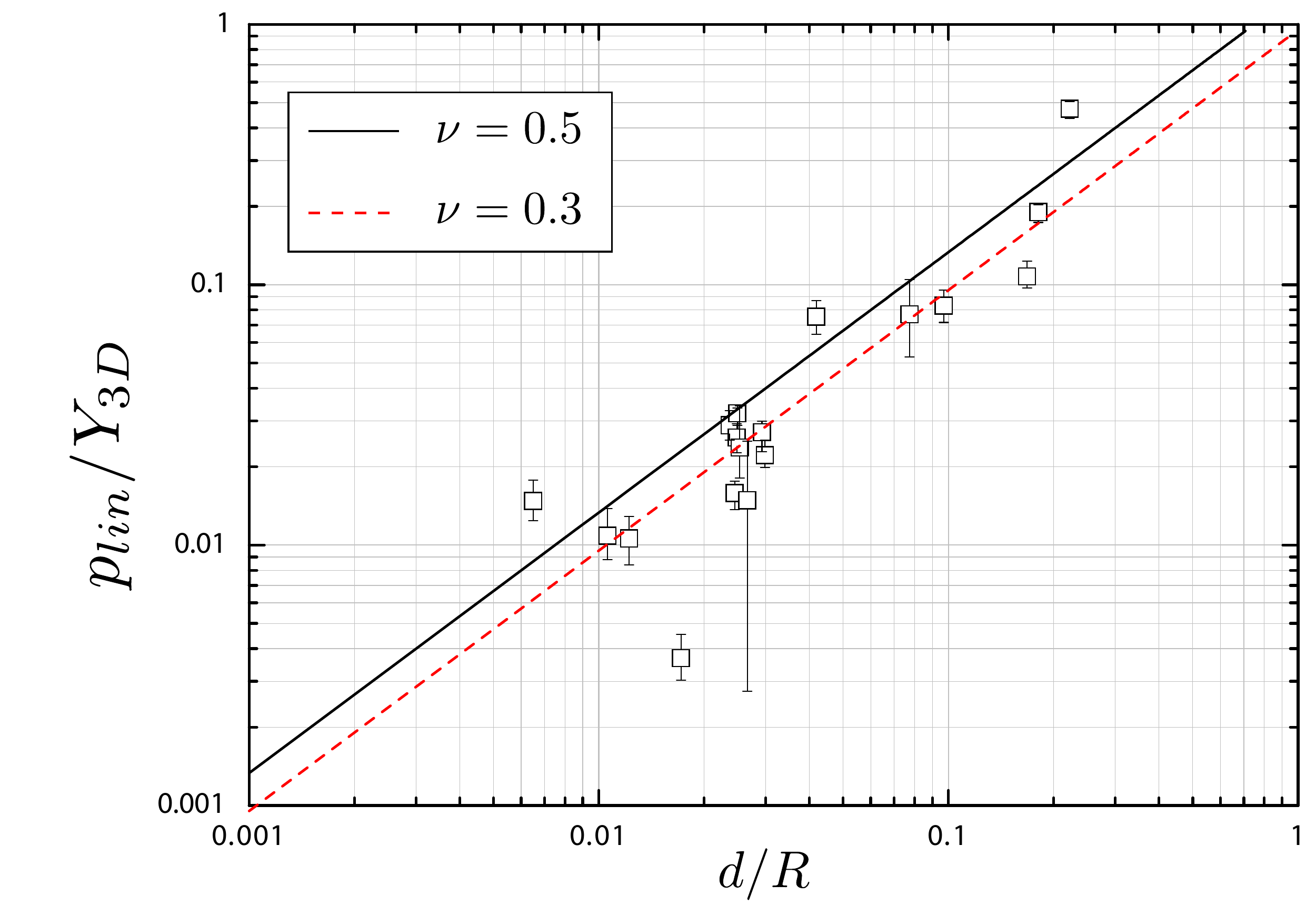}}

\caption{\label{fig:Traction}Slope $p_{lin}$ of the linear part of the $\Delta P\left(\frac{\Delta V}{V_{0}}\right)$
diagram, adimensionnalized by the Young modulus $Y_{3D}$ directly
measured from traction experiments. Straight lines : theoretical curves
for $p_{lin}/Y_{3D}=\frac{2}{3\left(1-\nu\right)}\times\frac{d}{R}$,
displayed for two values typically bounding the Poisson's ratio of
the shells studied.}
\end{figure}

Elastic properties (Young modulus and Poisson's ratio) were directly
measured with a tensile tester Shimadzu Autograph AGS-X machine equipped
with a 100 N load cell. The tensile tests were performed at ambient
temperature on dumbbell-shaped sample cut with a dogbone punch (gauge
length$18$\,mm $\times4$\,mm) in the ball, hence presenting a
thickness $d$. Traction was performed at a maximum crosshead speed
of 2 mm/min. For each ball, two different samples were submitted to
two tractions at a maximum deformation of 3\%, during which force
and elongations (both longitudinal and transversal, using instant
image treatment) were recorded. The true stress was plotted as a function
of the nominal strain, and the Young\textquoteright s modulus was
determined from the initial slope of the stress/strain curves. Video
recording of the sample during the deformation was performed in order
to measure the Poisson's ratio. Non-linearity between longitudinal and transversal
deformations prevented reliable measurement of $\nu$ for half of the samples. When we were able to unambiguously determine  its values, we found $0.45\le \nu\le 0.5$, as is typical for elastomeric materials.  Regardless, the Poisson's ratio has a small effect on values
of interest, as shown by theoretical curves of figures \ref{fig:Resum1}
and \ref{fig:Traction}. Figure \ref{fig:Traction} shows that there
is a satisfactory agreement for most of the shells between the slope  $p_{lin}$ of the $\wp(\Delta V)$ equilibrium
diagram in the isotropic deflation regime, adimensionalised by $Y_{3D}$ measured by traction experiments, and its theoretical value
computed from $d/R$ and $\nu$. We recall that most of the studied shells are low-cost toys obtained by rotational casting with  some variations of the thickness along the surface. These results indicates that for
moderate deformations, in-plane compression (that operates in deflation experiments)
and traction can be described using the same linear Young modulus.

\section{Conclusion and discussion}

Through theoretical and/or numerical studies, previous literature
provided hints about the behaviour of a ball that buckles under pressure,
according to its relative volume change, relative thickness and Poisson's
ratio. This was mostly obtained through the use of a model of elastic surface whose
range of validity is, a priori, restricted to thin shells ($d/R < 0.02$). The experimental
study conducted in this paper showed that thin shells deflate according to
these models, with quantitative agreement for the relationships between volume and inside-outside
pressure difference  controlled by the Young modulus of the ball. More surprisingly,
the agreement between the numerical deflation of elastic surfaces
and the experimental results on shells of an isotropic material still
holds for thicker shells (with important relative thicknesses, up
to almost  0.3), when the correspondence between 3D features and the
2D properties of the model surface is kept unchanged.

We also identified the dynamics for the rolling of the rim (which encloses
the depression formed during the buckling), with 1 or 2 relaxation
characteristic times, depending on the properties that are associated
to the dissipation in the material. We plan to run dynamical simulations in the future with models for shell membrane incorporating dissipation, so as to identify the source of these different times.

These results bring essential clues to the deflation of 
shells, and quantitative insights in a range of parameters that has
not yet been explored experimentally or theoretically.

\section{Acknowledgements}

We thank Pierre Saill\'e (CERMAV) for introducing us to traction experiments, and Guillaume Laurent and Antonin Borgnon for their involvement as students in the first experiments. A.D.'s position was funded by the European Research Council under the European UnionÕs Seventh Framework Programme (FP7/2007Ð2013)/ERC Grant No. 614655 Bubbleboost. 

\section{Authors contributions}
G.C. and C.Q. have designed the research and the experimental set-up. All the authors carried out the experiments. C.Q. has realised the additional numerical simulations.  G.C. and C.Q. were involved in the preparation of the manuscript.
All the authors have read and approved the final manuscript.

%
% BibTeX users please use
% \bibliographystyle{}
% \bibliography{}
%

%Author, \textit{Book title} (Publisher, place year) page numbers
% etc

\end{document}